\documentclass[prd,nofootinbib,preprint,superscriptaddress,secnumarabic]{revtex4}
\pdfoutput=1

\ifx\pdfoutput\undefined
\usepackage[dvips,bookmarks=false]{hyperref}	
\else
\usepackage{hyperref}	
\fi
\hypersetup{colorlinks,bookmarksopen,bookmarksnumbered,citecolor=blue,
linkcolor=blue,pdfstartview=FitH,urlcolor=blue}

\usepackage{graphicx}
\usepackage{amsmath}
\usepackage{amssymb}

\newcommand{\be}{\begin{equation}}
\newcommand{\ee}{\end{equation}}
\newcommand{\beq}{\begin{equation}}
\newcommand{\eeq}{\end{equation}}

\newcommand{\capdef}{}
\newcommand{\mycaption}[2][\capdef]{\renewcommand{\capdef}{#2}%
       \caption[#1]{{\footnotesize #2}}}
\makeatletter
\renewcommand{\fnum@table}{\textbf{\tablename~\thetable}}
\renewcommand{\fnum@figure}{\textbf{\figurename~\thefigure}}
\makeatother

\newcommand{\bs}[1]{{\ensuremath{\boldsymbol{#1}}}}

\begin{document}
\pagestyle{plain}

\vspace*{1cm}

\title{Dark Matter attempts for CoGeNT and DAMA
\vspace*{1cm}}

\def\mpi{Max-Planck-Institut f{\"u}r Kernphysik, PO Box 103980, 69029 Heidelberg, Germany}
\author{\textbf{Thomas Schwetz}\vspace*{3mm}}
\email{schwetz AT mpi-hd.mpg.de}
\affiliation{\mpi}

\def\Cincy{Department of Physics, University of Cincinnati, Cincinnati, Ohio 45221,USA}
\author{\textbf{Jure Zupan\footnote{On leave of absence from University of Ljubljana, Depart. of Mathematics and Physics, Jadranska 19, \\[-1mm]
 1000 Ljubljana, Slovenia and Josef Stefan Institute, Jamova 39, 1000 Ljubljana, Slovenia.}}\vspace*{3mm}}
\email{jure.zupan AT cern.ch}
\affiliation{\Cincy}


\begin{abstract}
\vspace*{5mm} 
Recently, the CoGeNT collaboration presented a positive signal for an annual modulation in their data set. In light of the long standing annual modulation signal in DAMA/LIBRA, we analyze the compatibility of both of these signal within the hypothesis of dark matter (DM) scattering on nuclei, taking into account existing experimental constraints. We consider the cases of elastic and inelastic scattering with either spin-dependent or spin-independent coupling to nucleons. We allow for isospin violating interactions as well as for light mediators. We find that there is some tension between the size of the modulation signal and the time-integrated event excess in CoGeNT, making it difficult to explain both simultaneously. Moreover, within the wide range of DM interaction models considered, we do not find a 
simultaneous explanation of CoGeNT and DAMA/LIBRA compatible with constraints from other experiments. 
However, in certain cases part of the data can be made consistent. For example, the modulation signal from CoGeNT becomes consistent with the total rate and with limits from other DM searches at 90~\%CL (but not with the DAMA/LIBRA signal) if DM scattering is inelastic spin-independent with just the right couplings to protons and neutrons to reduce the scattering rate on xenon. Conversely the DAMA/LIBRA signal (but not CoGeNT) can be explained by spin-dependent inelastic DM scattering.
\end{abstract}
\maketitle


\section{Introduction}

A number of experiments are searching with underground detectors for a signal from Dark Matter (DM) scattering on nuclei. A non-negligible interaction is for instance expected from DM that is a Weakly Interacting Massive Particle (WIMP). Two collaborations claim positive signals. Most recently, the CoGeNT experiment published results from 442 live days of data~\cite{Aalseth:2011wp}.
These confirm an event excess at low energies for the time integrated rate, already found in a previous release~\cite{Aalseth:2010vx}. In addition, the data show a $2.8\sigma$ indication in favor of an annual modulation of the event rate. In view of the long-standing
evidence for annual modulation reported by the DAMA/LIBRA
experiment~\cite{Bernabei:2008yi} (DAMA for short) an intriguing question
is, whether DM scattering can provide a consistent interpretation of the data from both
experiments. The relevant WIMP mass region of $m_\chi \sim 10$~GeV is also strongly constrained by
other experiments, e.g.,~\cite{Ahmed:2010wy, Angle:2011th, Aprile:2011hi}.
From a 4843~kg~day exposure of the XENON100 detector, three candidate events
have been observed, with a background expectation of $1.8\pm 0.6$
events~\cite{Aprile:2011hi}, showing no evidence for DM scattering. These
data lead not only to the world strongest limit for conventional WIMP
masses, constraining the DM--nucleon cross section to below $10^{-44} \,
{\rm cm}^2$ for $m_\chi \sim 100$~GeV, but have also important implications
for the low mass region around 10~GeV. 

Motivated by these recent developments we give a critical discussion of the CoGeNT data, and
reconsider possible DM interpretations of the global data, focusing on the
hints coming from CoGeNT and DAMA. We entertain the possibility that one DM species is responsible for the two signals and vary the properties of the interactions with nuclei. We consider spin-dependent and spin-independent interactions  for both elastic and inelastic scattering. We allow for general couplings to neutrons and protons, and we work in the contact interaction limit, but also analyze the effect of light mediators. Other analyses of the recent CoGeNT data have been performed in Refs.~\cite{Frandsen:2011ts,Hooper:2011hd, Belli:2011kw,Foot:2011pi} and we compare our results to those studies in case of overlap. For related analyses prior to new CoGeNT data see, e.g.,~\cite{Kopp:2009qt,Kopp:2010su,Kopp:2009et,Arina:2011si,Schwetz:2010gv,Fornengo:2010mk,Buckley:2010ve,Gunion:2010dy,Chang:2010en,Chang:2010pr,Savage:2010tg,Alves:2010pt,Graham:2010ca,Chang:2010yk,Andreas:2010dz,Fitzpatrick:2010em}. 

The outline of the paper is as follows. In sec.~\ref{sec:notations} we set the notation and define the DM interaction models used in the following. In sec.~\ref{sec:cogent} we consider the simplest case, namely spin-independent elastic scattering and argue that there is tension between the modulation signal and the integrated rate in CoGeNT data, and we show that within this framework CoGeNT and DAMA are incompatible and disfavored by other constraints. In secs.~\ref{sec:inel-iso} and \ref{sec:SD} we extend the analysis to inelastic scattering and generalized 
isospin dependence, for spin-independent and spin-dependent scattering, respectively. In sec.~\ref{sec:mediators} we consider the 
nontrivial energy dependence introduced by light mediator particles. We conclude in sec.~\ref{sec:conclusions}.

\section{Notation and definitions}
\label{sec:notations}

The differential cross section for non-relativistic DM scattering on a nucleon at rest, $\chi N\to \chi N$,  is ($N=n,p$) 
\be\label{eq:cs1}
\frac{d\sigma_N}{d \bs{q}^2} = \frac{1}{64 \pi}\frac{\overline{|\mathcal{M}|^2}}{ m_N^2 m_\chi^2 v^2} \,,
\ee
where $v$ is the velocity of $\chi$ and the momentum transfer, $\bs{q}^2$, is
related to the deposited energy $E_d$ as $\bs{q}^2 = 2 m_N E_d$. The nucleons are bound inside atomic nuclei with mass $m_A$, in which case the momentum exchange is $\bs{q}^2 = 2 m_A E_d$. In present-day DM experiments $|\bs{q}|\sim {\mathcal O}(20 {\rm ~MeV})-{\mathcal O}(100 {\rm ~MeV}).$ Assuming the same $E_d$ the corresponding $|\bs{q}|$ for scattering on a single $n$ or $p$ would be $|\bs{q}|\sim {\mathcal O}(0.2 {\rm ~MeV})-{\mathcal O}(1 {\rm ~MeV}).$

In the paper we will consider also the case of light mediators active in the DM--nucleon interaction. The assumed masses will be in MeV range or even lighter. For concreteness let us consider the case of DM that is a 
fermion, interacting with visible matter by emitting a scalar $\phi$ of mass $m_\phi$.  
The spin-averaged matrix element is given by
\be
\overline{|\mathcal{M}|^2} = 16 \, \lambda_\chi^2 \lambda_N^2 \frac{ \, m_\chi^2 m_N^2}
{(\bs{q}^2 + m_\phi^2)^2} \,,
\ee
where $\lambda_{\chi,N}$ are the couplings of the scalar to $\chi,N$,
\begin{equation}\label{Lint}
 {\cal L}_{\rm int}=\lambda_\chi \phi \bar \chi \chi +\sum_{N=p,n} \lambda_N \phi \bar N N.
\end{equation} 

The total cross section is obtained by integrating
eq.~\eqref{eq:cs1} from $\bs{q}^2 = 0$ to the maximum value allowed by
kinematics $\bs{q}^2_\mathrm{max} = 4\mu_{\chi N}^2 v^2$, with
$\mu_{ab} \equiv m_a m_b /(m_a + m_b)$. However, for a massless mediator
the total cross section diverges. Therefore we introduce a minimal momentum transfer
$\bs{q}^2_{\rm min}$ as an IR cut-off (corresponding to a threshold energy in an experiment) 
and define a reference cross section for DM--nucleon scattering as
\be
\sigma_N = \int_{\bs{q}^2_{\rm min}}^{4\mu_{\chi N}^2 v^2_{\rm ref}} 
d \bs{q}^2 \, \frac{d\sigma_N}{d \bs{q}^2} =
\frac{\lambda_\chi^2 \lambda_N^2}{\pi }\frac{ \mu_{\chi N}^2}
{\, (m_\phi^2 + \bs{q}^2_{\rm min})(m_\phi^2 + 4\mu_{\chi N}^2 v^2_{\rm
ref})} \,,
\ee
where we have neglected $\bs{q}^2_{\rm min}$ compared to $\bs{q}^2_{\rm
max}$ in the numerator.  Motivated by typical kinematics in DM direct detection experiment, we choose $\bs{q}^2_{\rm min} = (0.1 \, \rm MeV)^2$
and a reference velocity of $v_{\rm ref} = 10^{-3}\cdot c$. For $\mu_N \approx
m_N$ (which holds for $m_\chi \gg m_N \simeq 1$~GeV) this corresponds to 
$\bs{q}^2_{\rm max} \approx (2 \, \rm MeV)^2$.  Note that in the limit of
heavy mediators, $m_\phi^2 \gg \bs{q}^2_{\rm max}$, the cross
section becomes independent of $\bs{q}^2_{\rm min}$ and $v_{\rm ref}$, and $\sigma_N$ is the total cross section for $\chi N\to \chi N$ scattering, as usual.

The differential scattering rate 
in an experiment from $\chi$ scattering off a nucleus $(Z,A)$ 
in events/keV/kg/day is given by
\be
\frac{dR}{dE_d} = \frac{\rho_\chi}{m_\chi} \frac{1}{m_A} 
\int_{|\bs{v}| > v_{\rm min}} d^3v \frac{d\sigma_A}{d E_d} \, v f(v) \,,
\ee
where $\rho_\chi$ is the local DM density and $f(v)$ is the DM
velocity distribution in the rest frame of the detector. The lower limit of the integration is set by
the minimal velocity $v_{\rm min}$ that the incoming DM particle has
to have in order to be able to deposit an energy $E_d$ in the
detector. For the case of inelastic $\chi A\to \chi' A$ scattering it
is given by
\beq\label{vmin}
v_{\rm min}=\frac{1}{\sqrt{2m_A E_d}}\left(\frac{m_A
E_d}{\mu_{\chi A}}+\delta\right),
\eeq
where $\delta \equiv m_{\chi'} - m_\chi \ll m_\chi$. The same equation also
applies to elastic scattering, with $\delta=0$. The above example of
scalar interactions \eqref{Lint} leads to a spin independent (SI) scattering of DM on nuclei with the cross section 
given by
\be\label{eq:cs2}
\frac{d\sigma_A}{dE_d} = \frac{\lambda_\chi^2}{2\pi v^2} [Z \lambda_p F_p(\bs{q}^2) + (A-Z) \lambda_n F_n(\bs{q}^2) ]^2
\frac{m_A }{(\bs{q}^2 + m_\phi^2)^2}
\,,
\ee
where $\bs{q^2} = 2 m_A E_d$ and $F_{p,n}(\bs{q}^2)$ are nuclear form
factors. In general they can be different for neutrons and protons. In the
following we adopt the approximation $F_{n}(\bs{q}^2) =
F_{p}(\bs{q}^2) = F(\bs{q}^2)$, which corresponds to the assumption
that neutrons and protons (and consequently charge and mass) have the same distribution within the nucleus.  
Experimentally, the charge distributions within nuclei are well known from elastic electron scattering. Recently, reasonably precise information on mass distributions for some nuclei were also obtained from coherent photoproduction of $\pi^0$ mesons. Within ${\mathcal O}(5\%)-{\mathcal O}(10\%)$ experimental uncertainty the two distributions agree (see Table 3 of \cite{Krusche:2005jx}), while there is some preference for smaller neutron root mean square (rms) radii. In contrast, mean field estimates predict larger neutron distributions with 0.05\,fm--0.2\,fm bigger rms for the heavy nuclei \cite{Pomorski:1997fe}.
 
We use the Helm parameterization of the SI form factor, $F(\bs{q}^2) = 3 e^{-\bs{q}^2
s^2/2} [\sin(\kappa r)-\kappa r\cos(\kappa r)] / (\kappa r)^3$, with  $\kappa
\equiv |\bs{q}|$ and $s = 1$~fm, $r = \sqrt{R^2 - 5 s^2}$, $R = 1.2 A^{1/3}$~fm~\cite{Jungman:1995df}.  We have checked that the Lewin-Smith parameterization of $r,s$ \cite{Lewin:1995rx} gives compatible results, with the difference for instance for Xenon100 bounds being less than 3\% (see also the comparison in \cite{Duda:2006uk} of different form factor parameterizations and their effect on DM nucleon scattering). 

The coupling constants $\lambda_p$ and $\lambda_n$ are independent parameters of the model. If $\lambda_p=\lambda_n$ the DM--nucleon interactions are isospin conserving. We allow for general isospin breaking interactions, $\lambda_n \neq \lambda_p$, and define
\be\label{eq:sigmabarSI}
\tan\theta \equiv \frac{\lambda_n}{\lambda_p} \,,\qquad
\bar\sigma \equiv \frac{\sigma_n + \sigma_p}{2} \,.
\ee
Further we have to take into account that for a given target $Z$ several 
isotopes $A_i$ can be present. Neglecting the small effect of different 
isotopes on the kinematics and the form factors we define
\be\label{eq:Asq_eff}
A^2_{\rm eff} \equiv \sum_{i \in \rm isotopes} 2 r_i [Z\cos\theta + (A_i-Z)\sin\theta]^2 \,,
\ee
where $r_i$ are the relative abundances of the isotopes.
The SI cross section~\eqref{eq:cs2} can then be rewritten as
\be
\frac{d\sigma_A}{dE_d} = \frac{m_A \bar\sigma }{2 \mu_{\chi p} v^2} 
\, A_{\rm eff}^2 G(\bs{q}^2)F^2(\bs{q}^2) \,,
\ee
where we used $\mu_{\chi p} \approx \mu_{\chi n}$. The function $G(\bs{q}^2)$ 
takes into account the effect of a light mediator:
\be
G(\bs{q}^2) \equiv 
\frac{(m_\phi^2 + \bs{q}^2_{\rm min})(m_\phi^2 + 4\mu_{\chi N}^2 v^2_{\rm ref})}
{(m_\phi^2 + \bs{q}^2)^2}\,.
\ee
In the limit of heavy mediators, $m_\phi\gg 100$ MeV, this function goes to $G(\bs{q}^2) \to 1$.  If DM interactions are isospin conserving, $\lambda_p=\lambda_n$, the effective atomic weight is given by
$A_{\rm eff}^2 \to A^2 = \sum_i r_i A_i^2$, and one recovers the familiar expression for the SI cross section.

The SD scattering is treated in an analogous way, allowing for isospin violating interactions. For scattering on a nucleus
$A$ with spin $J$ we write
\be\label{eq:csSD}
\frac{d\sigma^{\rm SD}_A}{dE_d} = \frac{m_A}{2 \mu_{\chi p}^2 v^2} \,
\frac{4\pi \bar\sigma^{\rm SD}}{3 (2J+1)} G(\bs{q}^2) \,
[a_0^2 S_{00}(\bs{q}^2) + a_0 a_1 S_{01}(\bs{q}^2) + a_1^2 S_{11}(\bs{q}^2)]
\,,
\ee
with 
\be\label{eq:thetaSD}
\tan\theta \equiv \frac{a_n}{a_p} \,,\quad
a_0 \equiv \cos\theta + \sin\theta \,,\quad
a_1 \equiv \cos\theta - \sin\theta \,,\quad
\bar\sigma^{\rm SD} \equiv \sigma^{\rm SD}_n + \sigma^{\rm SD}_p \,,
\ee
$a_n,a_p$ are the couplings to the nucleons with $\sigma^{\rm SD}_{n,p}
\propto a_{n,p}^2$, and $S_{ij}$ are nuclear form factors. Note the slightly
different definition of $\bar\sigma^{\rm SD}$ compared to the SI case in
eq.~\eqref{eq:sigmabarSI}, in order to obtain the conventional normalization
in case of pure scattering off protons or neutrons. The SD form factors are
computed according to ref.~\cite{Toivanen:2009zz} for $^{133}$Cs (abundant
in the CsI crystals used by the KIMS experiment) and according to
ref.~\cite{Bednyakov:2006ux} for all other nuclei. 

In the analysis we fix the astrophysics
parameters and use everywhere a truncated Maxwellian velocity distribution in the galactic frame with escape velocity $v_{\rm esc}=550$ km/s, $v_0=230$ km/s and the local DM density $\rho_\chi=0.3 \, {\rm GeV}/{\rm cm}^3$. The impact of astrophysical uncertainties was discussed for example in~\cite{Fairbairn:2008gz,Lisanti:2010qx,Schwetz:2010gv} for elastic spin-independent scattering and for the inelastic case in~\cite{McCabe:2010zh}.

\section{Recent CoGeNT results and elastic SI scattering}
\label{sec:cogent}

We first discuss the recent  CoGeNT results~\cite{Aalseth:2011wp} in terms of
the most common DM models --- the ones leading to elastic scattering with spin-independent (SI) contact
interactions, assuming $f_n=f_p$ (this will be relaxed later). CoGeNT uses a germanium target with a very low threshold of 0.4~keVee (electron equivalent). To convert keVee into nuclear recoil energy measured in keV we use the quenching factor relation
$E [{\rm keVee}] = 0.199 (E [{\rm keV}])^{1.12}$~\cite{Barbeau:2007qi, collar}. Hence, the threshold of 0.4~keVee corresponds to a nuclear recoil energy of 1.9~keV, making CoGeNT especially sensitive to low mass WIMP scatterings.

\subsection{Modulation in CoGeNT?}

Let us evaluate the significance of the modulation signal reported in
CoGeNT~\cite{Aalseth:2011wp}, as well as whether a consistent
interpretation in terms of DM together with the unmodulated rate can be
obtained.

\begin{figure}
  \includegraphics[width=0.8\textwidth]{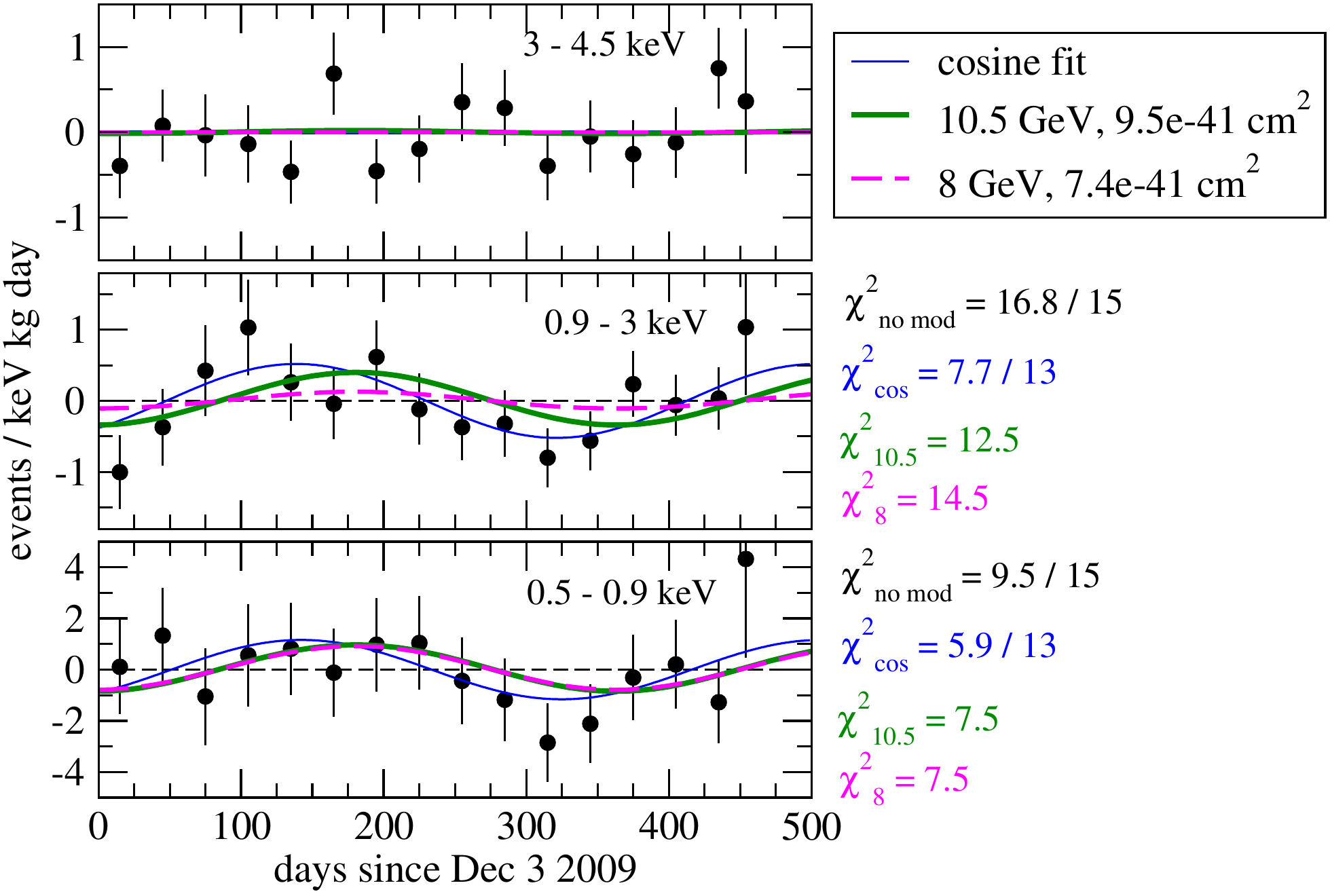}
  \mycaption{Time binned CoGeNT data~\cite{Aalseth:2011wp} with the mean
  values subtracted for
  three energy ranges. We show the results of fitting a cosine function with
  a period of one year and arbitrary phase (blue thin curve), the best fit
  to modulation data only (green thick curve) with $m_\chi = 10.5$~GeV, 
  and the best fit to the unmodulated data (magenta dashed curve) with
  $m_\chi = 8$~GeV. The corresponding $\chi^2$ values of the $0.5-0.9$ and
  $0.9-3$~keV bands are also shown.}
  \label{fig:cog-mod-data}
\end{figure}

Fig.~\ref{fig:cog-mod-data} shows the CoGeNT data in the energy bins
$0.5-0.9$~keV, $0.9-3$~keV, and $3-4.5$~keV. These data have been read off
from fig.~4 of \cite{Aalseth:2011wp}, where the $0.9-3$~keV band has been
obtained by subtracting the $0.5-0.9$~keV data from the $0.5-3$~keV data
shown there. First we note that in all three cases the goodness-of-fit of no
modulation is acceptable: we find $\chi^2$-values of 9.5, 16.8, 11.7 for
15~dof for the three bands (from low to high energies). However, if a cosine
with a period of one year, an amplitude $a$,
and a phase $t_0$ is fitted, the $\chi^2$ for the two
low energy bands improves significantly. We find
\begin{equation}
\begin{array}{l@{\quad}l@{\quad}l@{\quad}l}
  0.5-0.9\,{\rm keV}: & \Delta\chi^2_{\rm no\,mod} = 3.6\,, & 
    t_0 = 117 \pm 29 \,,& a = 1.24 \pm 0.65 \,,\\
  0.9-3  \,{\rm keV}: & \Delta\chi^2_{\rm no\,mod} = 9.1 \,,&
    t_0 = 109 \pm 18\,, & a = 0.56 \pm 0.18 \,,
\end{array}
\end{equation}
where $t_0$ is given in day of the year and $a$ in events/keV/day/kg. While
for the lowest energy band the significance for the modulation is modest, at
$1.9\sigma$, for the $0.9-3$~keV band we find slightly more than $3\sigma$
(these values have been obtained by evaluating the $\Delta\chi^2$ values for
1~dof). Note that the fitted $t_0$ is not in perfect agreement with the
expectation for a standard isotropic DM halo of $t_0 = 152$ (June $2^{\rm
nd}$), especially for the $0.9-3$~keV region, where it is about $2\sigma$
off.\footnote{A shift of the maximum of the DM scattering rate relative to June $2^{\rm nd}$ can occur in more complicated DM halos, see, e.g.,~\cite{Green:2003yh}.} For the highest energy region $3-4.5$~keV there is no significant preference for modulation.

\begin{figure}
  \includegraphics[height=0.44\textwidth]{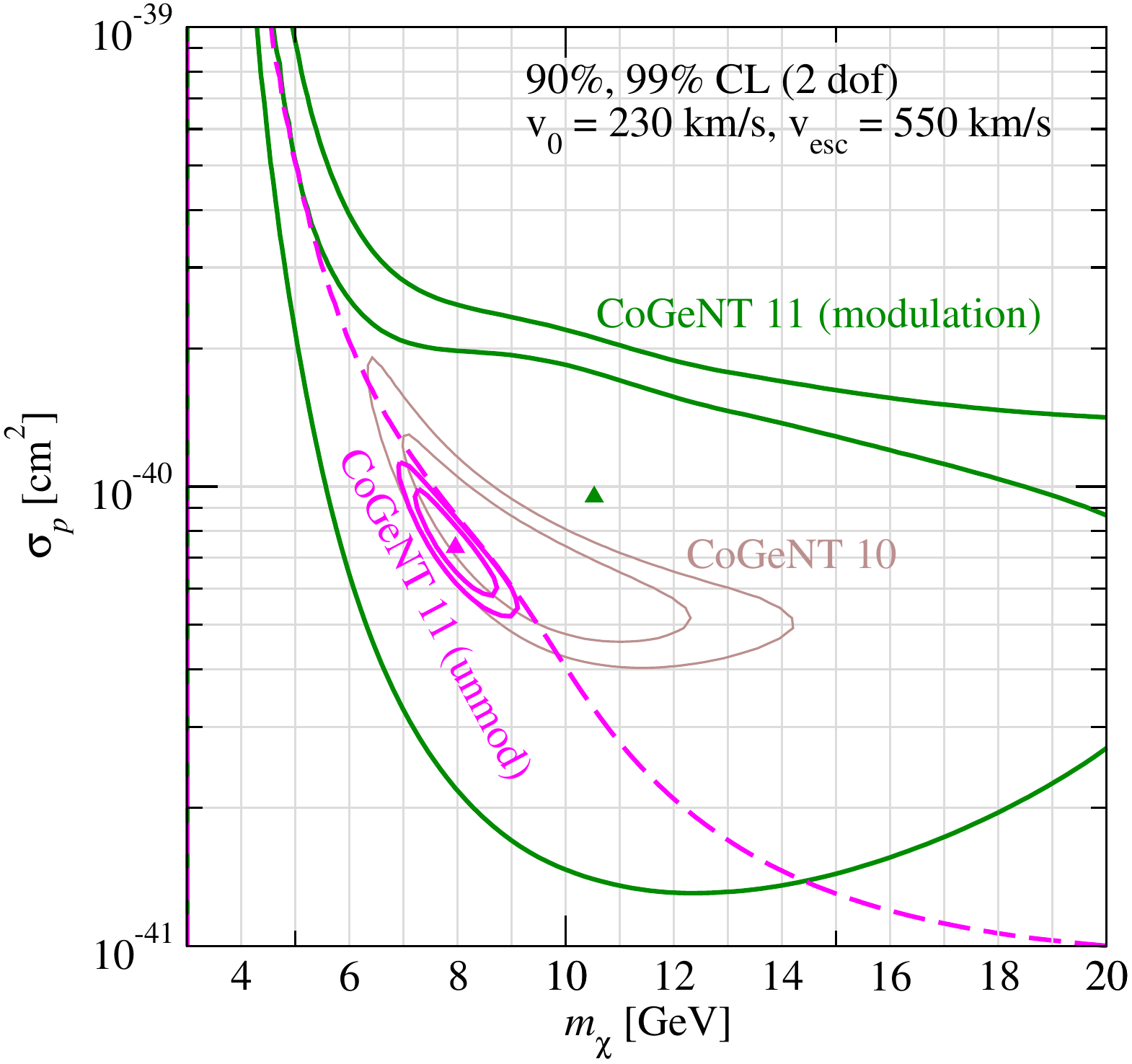} \quad
  \includegraphics[height=0.43\textwidth]{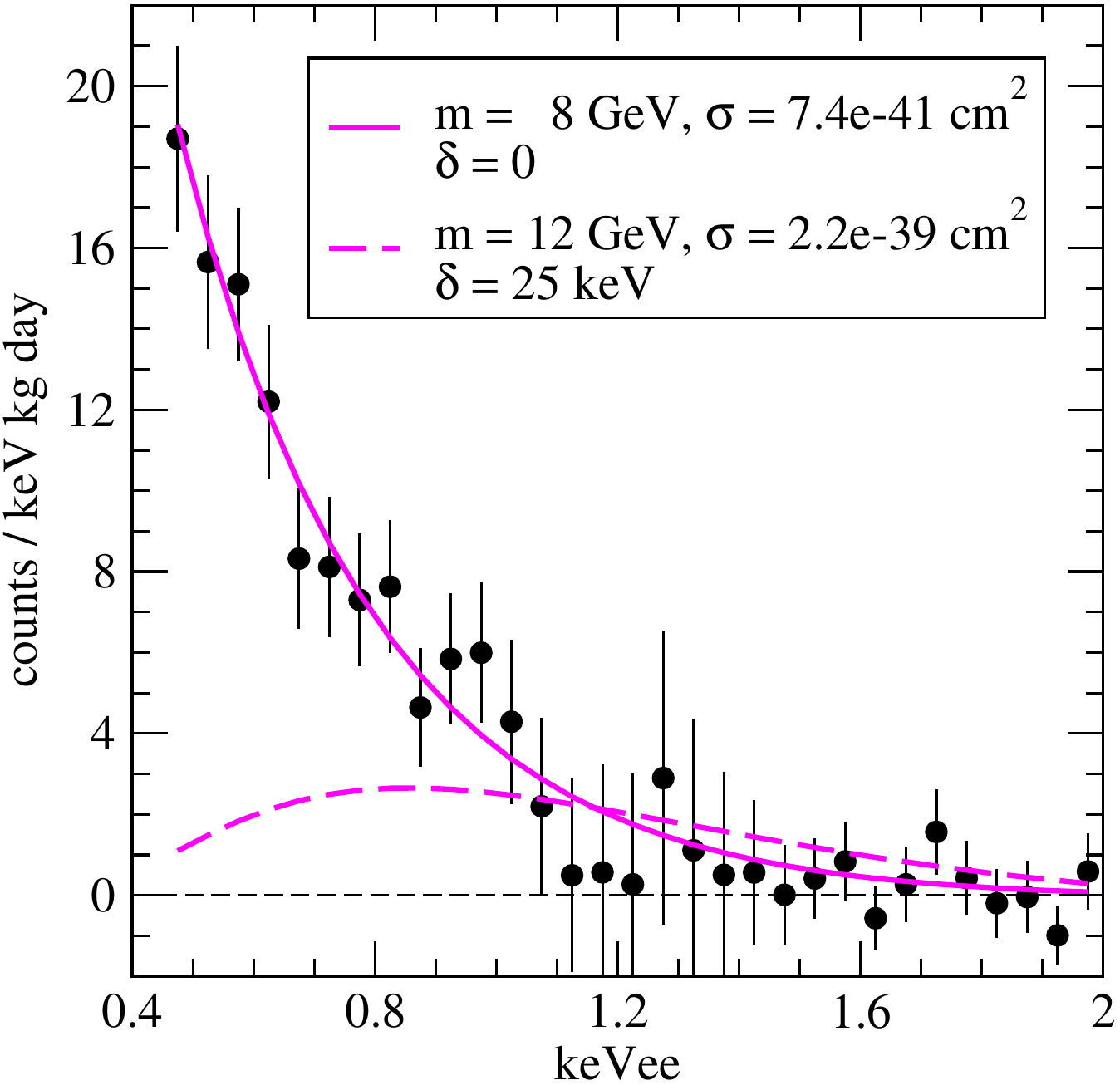}
  \mycaption{Left: regions in the DM mass--cross section plane for elastic
  spin-independent scattering at 90\% and 99\%~CL from CoGeNT
  data~\cite{Aalseth:2011wp} using only the modulation data (green) or only
  the unmodulated rate (magenta). The corresponding best fit points are
  marked with triangles. For the fit to the unmodulated rate we assume that
  the total excess events (after subtraction of the 
  L-shell peaks and a flat
  background) are explained by DM.  The dashed curve corresponds to the
  99\%~CL limit from CoGeNT unmodulated data requiring that not more events
  than observed are predicted. We also show the allowed regions from CoGeNT
  2010 data~\cite{Aalseth:2010vx}. Right: unmodulated CoGeNT event spectrum
  and two predictions from DM. The solid (dashed) curve refers to elastic
  (inelastic) scattering with parameters as given in the legend.}
  \label{fig:cog-regions}
\end{figure}

Now we proceed to a fit of these data in terms of elastic spin-independent
DM scattering. We find a best fit point with $m_\chi = 10.5$~GeV and
$\sigma_p = 9.5\times 10^{-41}\,{\rm cm}^2$. The corresponding prediction
for the modulation is shown in fig.~\ref{fig:cog-mod-data} as a thick green curve.
We observe the shift in the phase compared to the free cosine fit which
leads to a worsening of the fit: the best fit DM predictions have only
$\Delta\chi^2 = 2 \, (4.3)$ for $0.5-0.9\, (0.9-3)$~keV compared to no
modulation. The fit regions in the plane of DM mass and cross section are
shown in fig.~\ref{fig:cog-regions}. The reduced $\Delta\chi^2$ values of
the best fit point compared to no modulation imply that a closed region
appears only at 90\%~CL, whereas at 99\%~CL data is compatible with no
signal. 

In fig.~\ref{fig:cog-regions} we show also the allowed region obtained from
the unmodulated event rate. We use the data shown in the in-set of fig.~1 in
\cite{Aalseth:2011wp}, reproduced in the right panel. This is the
unexplained event excess which remains after subtracting the 
L-shell peaks
as well as a flat background. In the plot of \cite{Aalseth:2011wp} data
points corresponding to two different peak-subtraction methods are shown
(black and white data points). For our fit we take the average of the two
points and conservatively use the lowest and highest edges of the error bars
to account for this systematic uncertainty in the fit. For the region
delimited by the magenta ellipses in fig.~\ref{fig:cog-regions} we assume
that the total event excess is explained by DM events. In this case the best
fit point is obtained at $m_\chi = 8$~GeV and $\sigma_p = 7.4\times
10^{-41}\,{\rm cm}^2$, corresponding to the solid curve in the right panel. 
The dashed curve in the left panel shows the limit which is obtained by
just requiring that not more events are predicted from DM than observed.

We see from fig.~\ref{fig:cog-regions} that the best fit point from the
modulation signal is excluded by the unmodulated rate. The allowed region
from the unmodulated spectrum is still located within the 90\%~CL region of
the modulated data, however, for those parameter values the modulation in
the $0.9-3$~keV region is significantly reduced. This is shown as dashed
magenta curves in fig.~\ref{fig:cog-mod-data}. While the prediction in the
lowest energy band is essentially identical to the one of the modulation
best fit point, the modulation amplitude is reduced significantly for 
$0.9-3$~keV due to the steeper shape of the energy spectrum at lower DM
mass. The $\Delta\chi^2$ compared to no modulation is now only 2.3, which
means that the modulation is actually not explained in the region where the
signal is strongest. 

A similar result is obtained for all parameter points to the left of the
dashed curve in fig.~\ref{fig:cog-regions}. We conclude that in the simplest
case of elastic spin-independent scattering within a standard halo there is tension between the
modulation signal and the unmodulated spectrum in CoGeNT.

\subsection{CoGeNT vs DAMA vs CDMS vs XENON100}

\begin{figure}
  \includegraphics[width=0.5\textwidth]{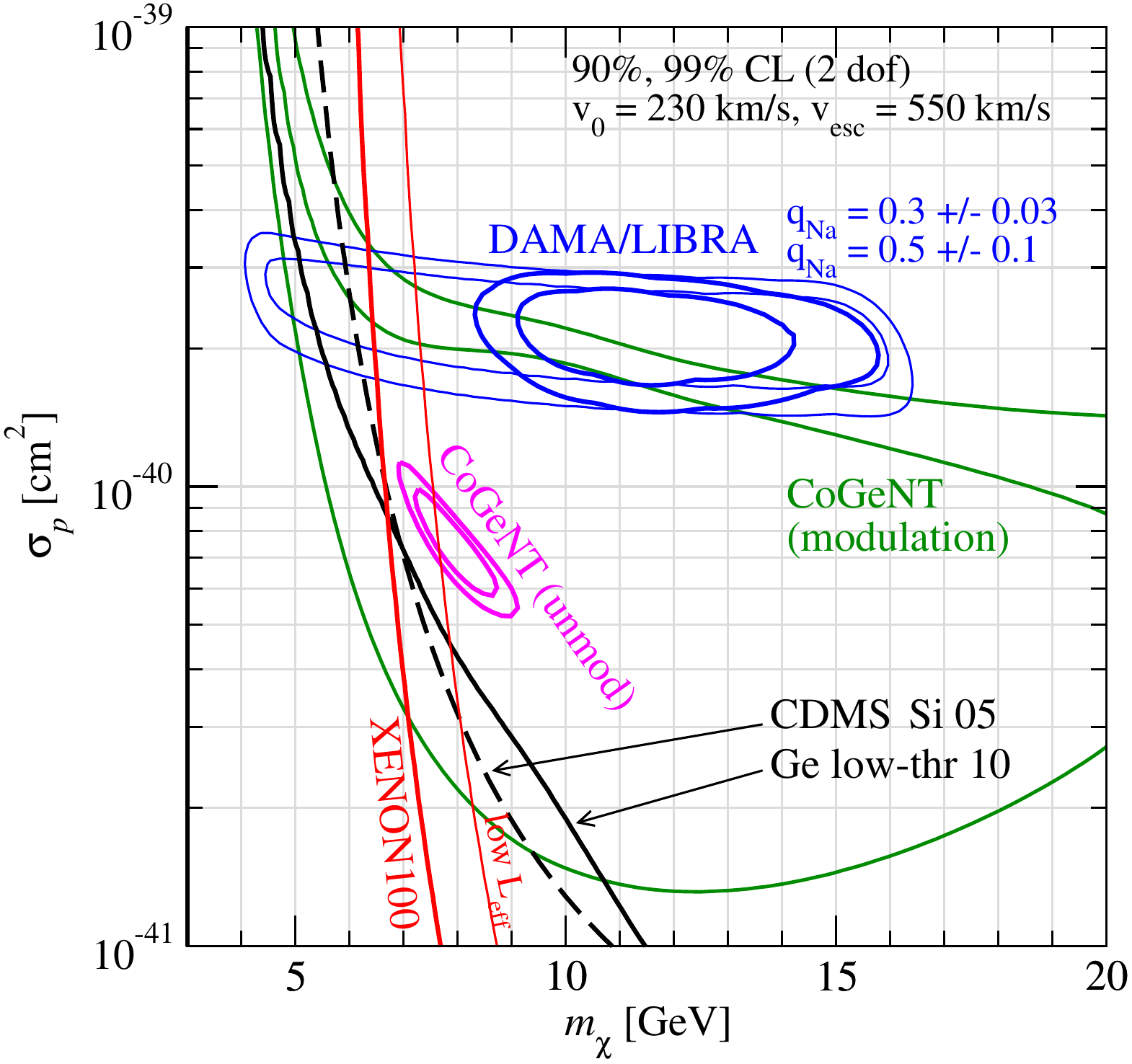}
  \mycaption{Allowed regions (90\% and 99\%~CL) and limits (90\%~CL) in the
  DM mass--cross section plane for elastic spin-independent scattering. We
  use data from CoGeNT~\cite{Aalseth:2011wp} (modulated and unmodulated),
  DAMA/LIBRA~\cite{Bernabei:2008yi} modulation, CDMS low-threshold Ge
  \cite{Ahmed:2010wy}, CDMS Si~\cite{Akerib:2005kh}, and
  XENON100~\cite{Aprile:2011hi}.  For DAMA/LIBRA we show the allowed regions
  for two illustrative assumptions on the quenching factor of sodium, for
  XENON100 we show the limit for two assumptions on the light-yield factor
  $L_{\rm eff}$.} \label{fig:regions-all-eSI}
\end{figure}

Let us now discuss whether the CoGeNT results can be reconciled with other
data in the framework of elastic spin-independent scattering. First we
consider the DAMA/LIBRA data~\cite{Bernabei:2008yi}, which show annual
modulation at significance of about $8.9\sigma$. We fit the lowest 12 bins in energy 
for the modulated event rate in DAMA, and use the unmodulated spectrum as an upper bound on the predicted rate, see~\cite{Fairbairn:2008gz, Kopp:2009qt} for details. In the following we
assume that ion-channeling does not occur, following the results
of~\cite{Bozorgnia:2010xy}. This leaves scattering on the Na atoms as
possible explanation of the DAMA signal~\cite{Bottino:2003cz, Gondolo:2005hh}. Hence the
quenching factor of sodium, $q_{\rm Na}$, is an important input into the
analysis. In our fit we take into account the uncertainty of $q_{\rm Na}$
and marginalize the $\chi^2$ with respect to this parameters. In
fig.~\ref{fig:regions-all-eSI} we show the DAMA allowed region for the two
assumptions $q_{\rm Na} = 0.3\pm 0.03$ (the ``default'' value used in many
analyses) and $q_{\rm Na} = 0.5\pm 0.1$. We have arbitrarily adopted such a
high value, following the suggestion of \cite{Hooper:2010uy} to reconcile
CoGeNT and DAMA regions. We stress, however, that a recent measurement in
\cite{Chagani:2008in} obtains a value of $q_{\rm Na} = 0.25\pm 0.06$ at
$E_{\rm nr} = 10$~keV, which taken at face value strongly disfavors large
quenching factors. In particular, the example of $q_{\rm Na} = 0.5\pm 0.1$
used in fig.~\ref{fig:regions-all-eSI} is in clear conflict with the
measurement from \cite{Chagani:2008in} and we adopt it only for illustrative
purpose. Most of the older measurements compiled in Tab.~10 of \cite{Bernabei:2003za} also seem to indicate a value $q_{\rm Na} \simeq 0.3$. The DAMA allowed region is clearly separated from the
region which can explain the CoGeNT unmodulated event excess, while there is some overlap of the region indicated by  the CoGeNT modulation data.  This holds
even for the extreme assumption on the sodium quenching factor.\footnote{Let us mention that our results differ here from the ones of Ref.~\cite{Hooper:2011hd} where a better agreement between CoGeNT and DAMA is obtained. The region from the CoGeNT unmodulated event
rate in \cite{Hooper:2011hd} is obtained at somewhat larger cross sections, whereas the DAMA region is found at cross sections of about a factor two lower than ours. Our DAMA region agrees, e.g., with the one of Ref.~\cite{Savage:2010tg}.}

Furthermore, fig.~\ref{fig:regions-all-eSI} shows that severe constraints
from CDMS and XENON100 disfavor the region indicated by both, CoGeNT and
DAMA. From CDMS we show in the figure limits from a dedicated low-threshold
analysis of CDMS-II Ge data~\cite{Ahmed:2010wy} using data down to a
threshold of 2~keV, as well as data from a somewhat old data sample on
Si~\cite{Akerib:2005kh} of 12~kg~day exposure with a threshold at 7~keV. As
discussed in \cite{Collar:2011kf}, the energy scale determination is crucial
for these limits. We have estimated that an ad-hoc shift in the energy scale
of 40\% may render the limit from \cite{Ahmed:2010wy} consistent with the
CoGeNT region. Let us stress that we do not advocate such a shift, and the
critique of \cite{Collar:2011kf} has been addressed in great detail in the
appendix of \cite{Ahmed:2010wy} available in the arXiv version of that
paper. 

XENON100 published results from a 4843~kg~day exposure~\cite{Aprile:2011hi},
obtaining three candidate events with a background expectation of $1.8\pm
0.6$ events. We derive a conservative limit by using the maximum gap method
from~\cite{Yellin:2002xd}. An important issue in the interpretation of data
from xenon experiments is the scintillation light yield, conventionally
parametrized by the function $L_{\rm eff}(E_{\rm nr})$, see, e.g.,~\cite{Savage:2010tg}.
An improved measurement of this function has been published
recently~\cite{Plante:2011hw}. For the XENON100 limit (thick curve) in
fig.~\ref{fig:regions-all-eSI} we use for $L_{\rm eff}(E_{\rm nr})$ the
black solid curve from fig.~1 of~\cite{Aprile:2011hi}, which has been
obtained from a fit to various data, dominated by the 
measurement~\cite{Plante:2011hw} at low energies. Such values of $L_{\rm eff}$ are also supported by the analysis of Ref.~\cite{Bezrukov:2010qa} by modeling a relation between scintillation and ionization and 
using existing low energy data on ionization. In order to estimate the
impact of the uncertainties related to the scintillation light yield we show
also the limit (thin curve) which we obtain by using the lower edge of the
light blue region in fig.~1 of~\cite{Aprile:2011hi} corresponding to the
lower $2\sigma$ range for $L_{\rm eff}$. Nevertheless, we have also checked that the
assumptions on $L_{\rm eff}$ below $E_{\rm nr} \approx 3$~keV (where no data
are availble) have no impact on the XENON100 curve, and the limit remains
unchanged even assuming $L_{\rm eff} = 0$ below 3~keV. Hence, the result
does not rely on any extrapolation into a region with no data. 

Fig.~\ref{fig:regions-all-eSI} shows that the XENON100 limit using the
best-fit curve for $L_{\rm eff}$ excludes the CoGeNT (unmod) favored region,
while pushing $L_{\rm eff}$ down by 2$\sigma$ leads to a region marginally
consistent. In both cases large values for $q_{\rm Na}$ have to be assumed
to make DAMA consistent with the limits. Let us note that a dedicated
analysis of XENON10 data~\cite{Angle:2011th} obtains the energy scale from
the ionization signal and is therefore independent of the scintillation
efficiency $L_{\rm eff}$. These results put additional severe constraints on
the region of interest and exclude both the regions indicated by DAMA and
CoGeNT (not shown).

\section{Inelastic scattering and nontrivial isospin dependence}
\label{sec:inel-iso}

In the previous section we have seen that it is difficult to obtain a
consistent interpretation of all data in terms of elastic spin-independent
interactions. Here we investigate some possibilities beyond this simplest
mechanism of DM interactions. 

\subsection{Inelastic scattering}
\label{sec:inel}

We start by considering inelastic scattering. This has been proposed in
\cite{TuckerSmith:2001hy} in order to reconcile the DAMA modulation signal
with the other constraints. For inelastic scattering the modulation signal is
enhanced compared to the unmodulated rate and in general heavy target nuclei
are favored. While recent data from XENON100~\cite{Aprile:2011ts} and
CRESST-II (tungsten)~\cite{seidel_iDM10} disfavor such an interpretation of
DAMA, it might be interesting to use inelastic scattering to resolve the
tension between the modulated and unmodulated rate in
CoGeNT~\cite{Frandsen:2011ts}.

\begin{figure}
  \includegraphics[width=0.5\textwidth]{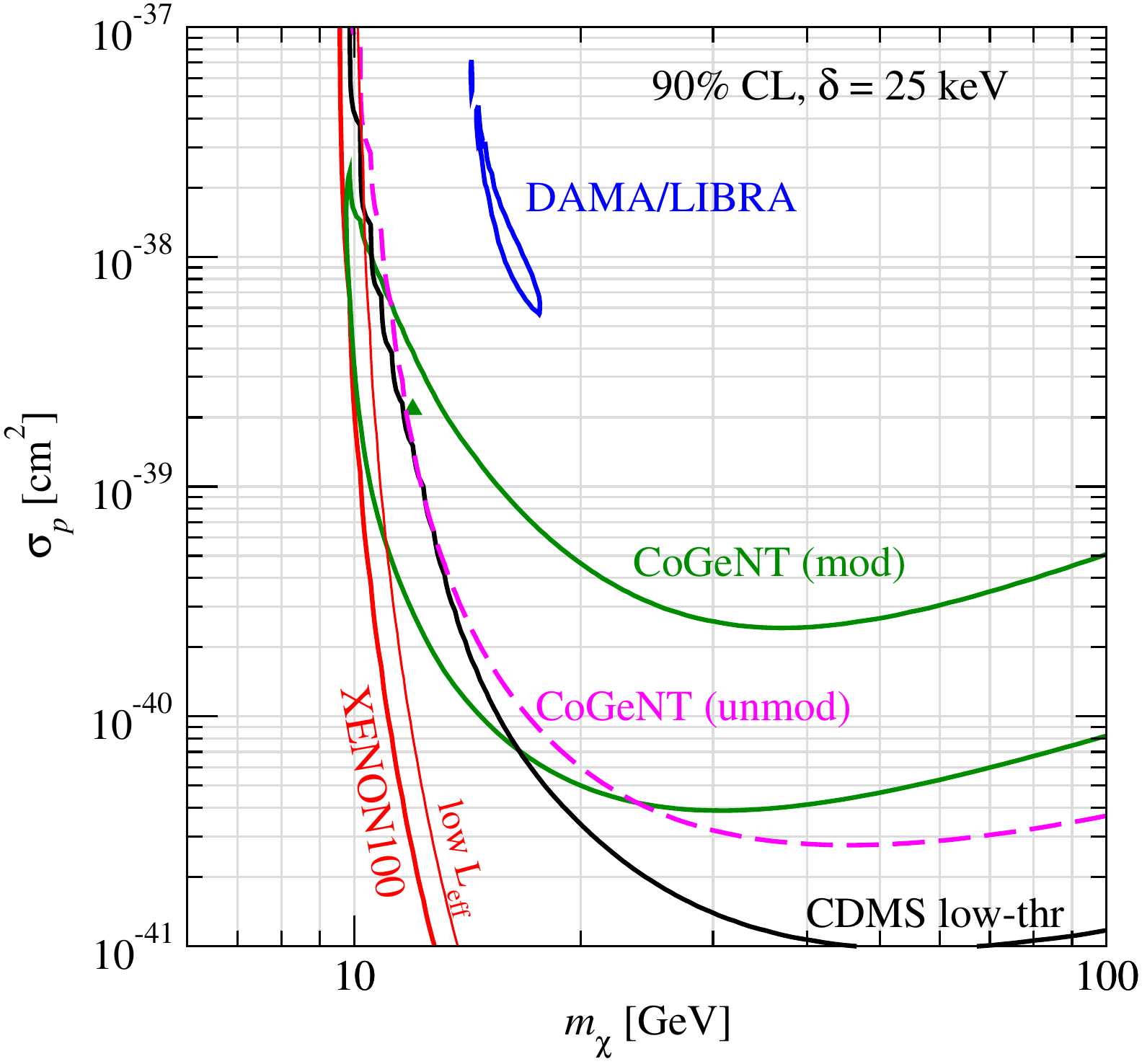}
  \mycaption{Allowed regions and limits at 90\%~CL in the
  DM mass--cross section plane for inelastic spin-independent scattering
  with a DM mass spitting $\delta= 25$~keV. We
  use data from CoGeNT~\cite{Aalseth:2011wp} (modulated and unmodulated),
  DAMA/LIBRA~\cite{Bernabei:2008yi}, CDMS low-threshold Ge \cite{Ahmed:2010wy}, and
  XENON100~\cite{Aprile:2011hi}.} \label{fig:cogent-iSI}
\end{figure}  

In fig.~\ref{fig:cogent-iSI} we show the result for a DM mass splitting of
$\delta = 25$~keV, which gives the best compatibility of modulated CoGeNT
data with the unmodulated rate. As visible in the plot the best fit point
for the modulation, $m_\chi = 12$~GeV, $\sigma_p = 2.2\times
10^{-39}\,{\rm cm}^2$, is obtained just at the 90\%~CL limit of the
unmodulated spectrum (used as an upper limit) as well as the CDMS Ge
low-threshold bound. The predicted spectrum for the modulation as well as
the $\chi^2$ is practically the same as for the elastic best fit point at
$m_\chi = 10.5$~GeV shown in fig.~\ref{fig:cog-mod-data}. Note,
however, that inelastic scattering does not offer an explanation for the
unmodulated event excess, as shown by the dashed curve in
fig.~\ref{fig:cog-regions} (right). The modified kinematics lead to a
different shape of the spectrum which cannot explain the exponential event
excess towards low energies.

From fig.~\ref{fig:cogent-iSI} we observe also that there is a clear mismatch
between the DAMA and CoGeNT regions. Furthermore, the conflict with the
XENON100 bounds persists, even in the case of using  
$L_{\rm eff}$ at the
lower $2\sigma$ limit. If the parameter space to the left of the XENON100
curve is considered, we find that again the modulation in the $0.9-3$~keV
CoGeNT data cannot be explained. We conclude that inelastic scattering
does not provide a satisfactory description of all data.

\subsection{Generalized isospin dependence}

Motivated by the observation that Higgs-mediated SI scattering leads to
roughly equal couplings of DM to neutron and proton (due to the dominance of
the strange quark contribution), conventionally $f_n\approx f_p$ is assumed,
which leads to an $A^2$ factor for the SI cross section. This assumption
needs not to be fulfilled in general. As pointed out recently in
\cite{Chang:2010yk, Feng:2011vu} there might be negative interference
between scattering off neutrons and protons, which would lead to a vanishing
cross section for $f_n/f_p = -Z/(A-Z)$ 
(for a realization of such isospin violating DM in a technicolor model see
\cite{DelNobile:2011je}). If a given element consists only of
one isotope the cancellation can be complete, whereas if different isotopes
are present only the contribution of one particular isotope can be cancelled
exactly. 

\begin{figure}
  \includegraphics[height=0.44\textwidth]{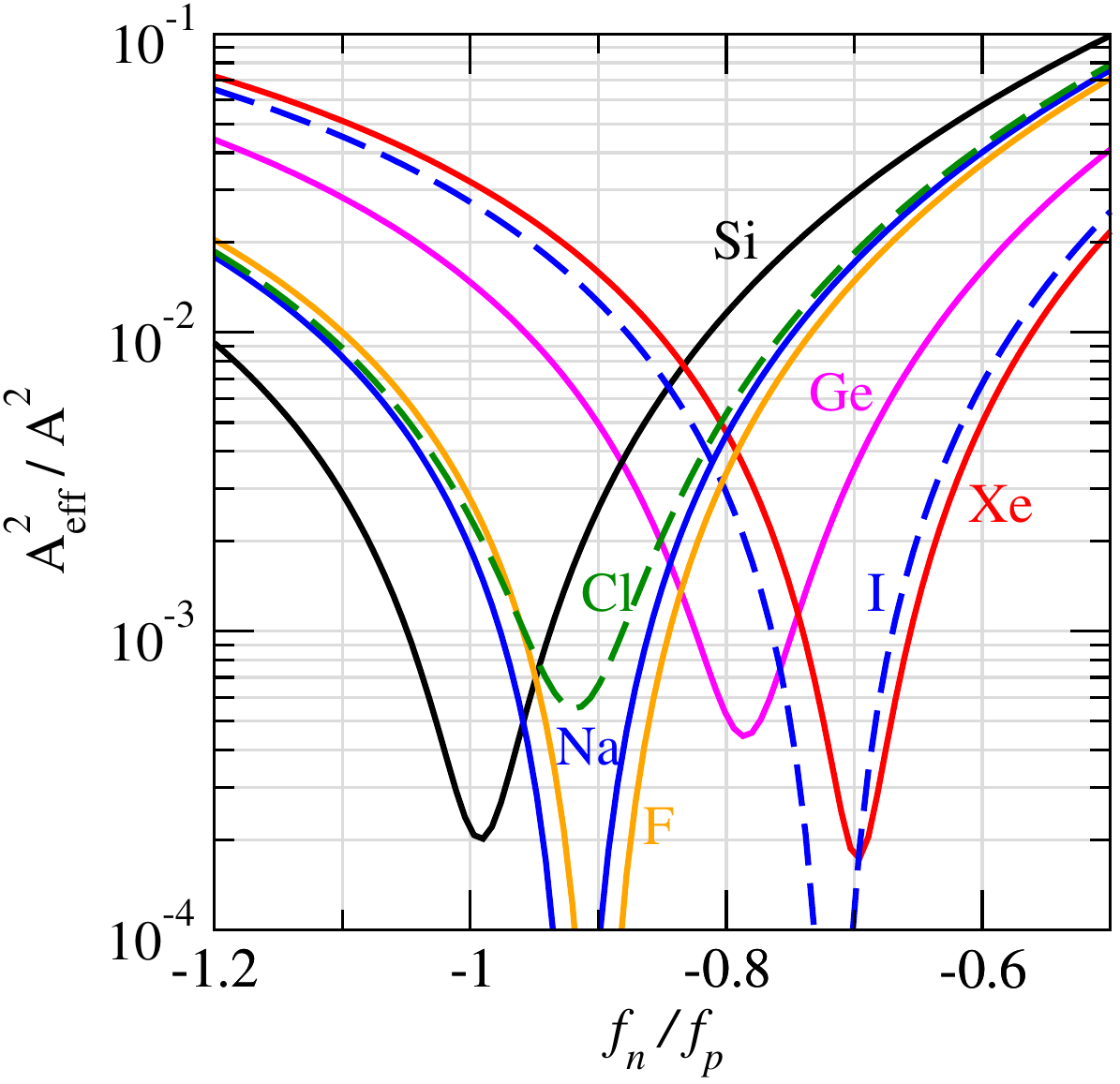} \quad
  \includegraphics[height=0.44\textwidth]{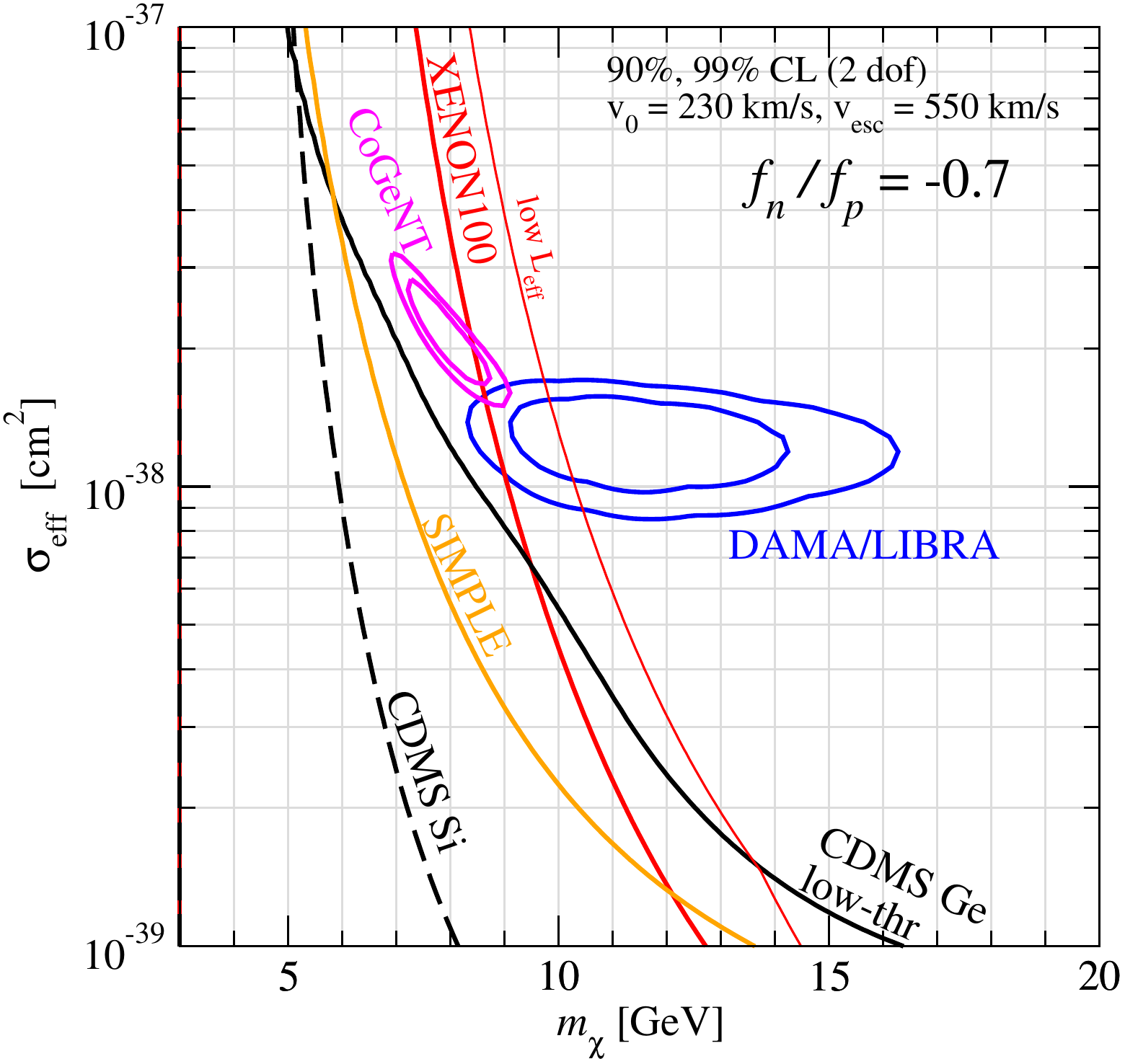}
  \mycaption{Left: $A^2_{\rm eff}/A^2$ defined in eq.~\eqref{eq:Asq_eff}, as a
  function of the ratio of DM coupling to neutron and proton for various
  elements. Right: regions (90\% and 99\%~CL) and limits (90\%~CL) in the DM
  mass--cross section plane for elastic spin-independent scattering with
  $f_n/f_p = -0.7$. The cross section on the vertical axis corresponds to $\bar\sigma$ defined in eq.~\ref{eq:sigmabarSI}. We use data from CoGeNT~\cite{Aalseth:2011wp}
  (unmodulated spectrum), DAMA/LIBRA~\cite{Bernabei:2008yi}, CDMS Ge
  low-threshold \cite{Ahmed:2010wy} and Si~\cite{Akerib:2005kh}, 
  XENON100~\cite{Aprile:2011hi}, and SIMPLE~\cite{Felizardo:2011uw}.}
  \label{fig:isospin}
\end{figure}

Assuming natural isotopic abundances we show in the left panel of
fig.~\ref{fig:isospin} the suppression factor compared to the case $f_n =
f_p$, where the effective atomic number in case of general couplings,
$A^2_{\rm eff}$, is defined in eq.~\eqref{eq:Asq_eff}. We zoom into the region
around the neutron/proton cancellation $f_n / f_p \approx -1$. One
observes that due to the different neutron/proton ratios in the various
elements, the cancellations occur at different values of $f_n / f_p$ for
different elements, which might allow better compatibility of some
experiments. In particular, there is a minimum of the cross section for
xenon at $f_n/f_p \approx -0.7$, which can be used to weaken the limit from
XENON100 compared to CoGeNT (Ge) by more than one order of magnitude in the
cross section \cite{Chang:2010yk, Feng:2011vu, Frandsen:2011ts}.

Fig.~\ref{fig:isospin} (right) shows the DAMA and CoGeNT allowed regions compared to limits from various other experiments for $f_n/f_p = -0.7$, chosen in order to weaken the XENON100 constraint. Due to the enhancement of Na compared to Ge (see left panel) the DAMA region appears at lower cross sections and overlaps with the CoGeNT region at 99\%~CL, and indeed consistency with XENON100 is obtained, especially if the uncertainty on $L_{\rm eff}$ is taken into account. Obviously, the CDMS low-threshold constraint from Ge cannot be circumvented due to the same material as in CoGeNT. Moreover, scattering on Si is enhanced compared to Ge (see left panel), which makes the conflict with the CDMS Si data more severe. In addition we show in this plot also a limit from the SIMPLE experiment~\cite{Felizardo:2011uw}. This experiment uses a C$_2$ClF$_5$ target and upper limits of 0.289 and 0.343 events/kg/day at 90\%~CL have been reported from 13.47 and 6.71~kg~day exposures, respectively, with a threshold of 8~keV. The limits from SIMPLE are less stringent in the cases  discussed so-far, but as visible from the left panel of fig.~\ref{fig:isospin} the scattering on both Cl and F is enhanced compared to Ge for $f_n/f_p = -0.7$, and in this case SIMPLE excludes the CoGeNT/DAMA region.

We conclude that invoking a cancellation between neutrons and protons can lead to a consistent description of DAMA and CoGeNT (unmodulated excess), and weaken the XENON100 limit sufficiently. However, the corresponding parameter region is excluded by CDMS Ge, Si and SIMPLE data. While we have shown results explicitly only for the choice $f_n/f_p = -0.7$ which minimizes the cross section for xenon, it is clear from fig.~\ref{fig:isospin} (left) that also other choices for $f_n/f_p$ cannot improve the global situation.

\subsection{Combining inelasticity and general isospin couplings}

\begin{figure}
  \includegraphics[height=0.44\textwidth]{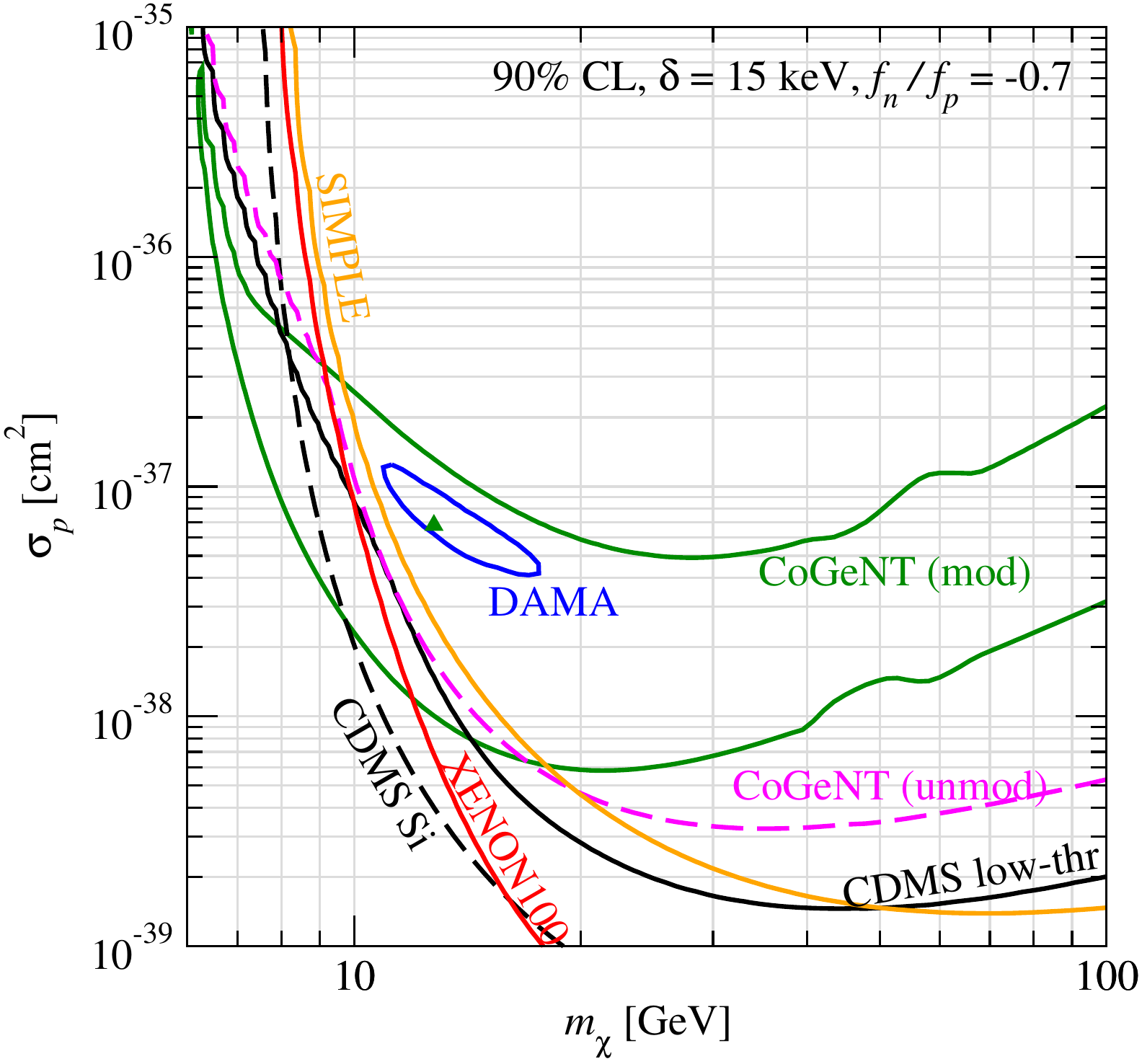} \quad
  \includegraphics[height=0.44\textwidth]{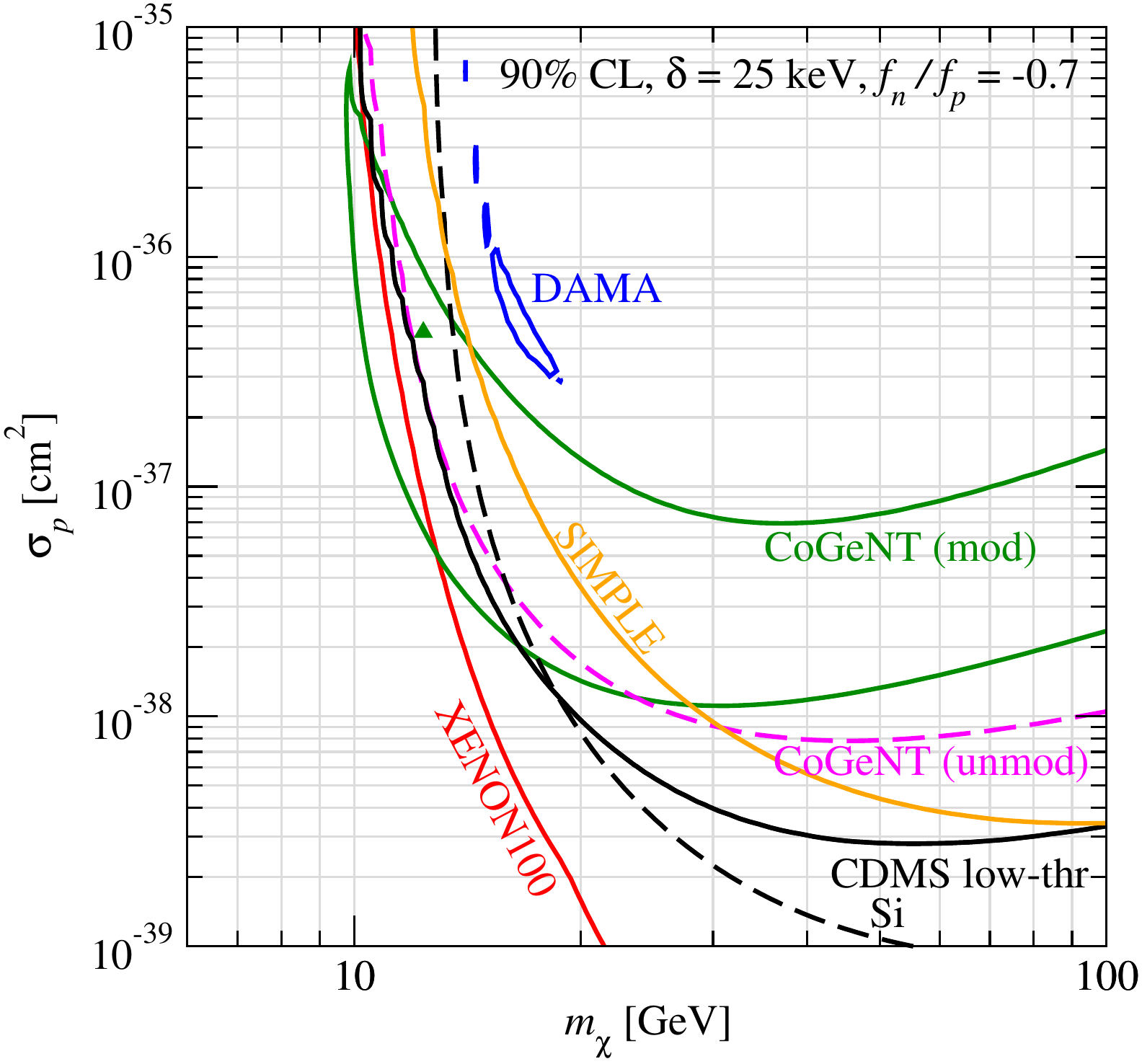}
  \mycaption{Allowed regions and limits at 90\%~CL in the DM
  mass--cross section plane for inelastic spin-independent scattering with
  $f_n/f_p = -0.7$ and DM mass splittings $\delta = 15$~keV (left) and $\delta = 25$~keV (right).
  The cross section on the vertical axis corresponds to $\bar\sigma$ defined in eq.~\ref{eq:sigmabarSI}. We use data from CoGeNT~\cite{Aalseth:2011wp}
  (modulation and limit from the unmodulated spectrum), DAMA/LIBRA~\cite{Bernabei:2008yi}, CDMS Ge
  low-threshold \cite{Ahmed:2010wy} and Si~\cite{Akerib:2005kh}, 
  XENON100~\cite{Aprile:2011hi}, and SIMPLE~\cite{Felizardo:2011uw}. The best fit point of the CoGeNT modulation data is marked with a triangle.}
  \label{fig:isospin-inel}
\end{figure}  

It has been suggested in \cite{Frandsen:2011ts} to use inelastic scattering
as well as $f_n/f_p=-0.7$ in order to reconcile the CoGeNT modulation with
DAMA and other constraints. We 
reconsider this idea in
fig.~\ref{fig:isospin-inel}, showing regions and limits for DM mass
splittings of $\delta = 15$ and 25~keV. As discussed in sec.~\ref{sec:inel},
for inelastic scattering the unmodulated spectrum in CoGeNT cannot be fit
due to the specific spectral shape of the signal. Therefore we use the
unmodulated rate only as an upper limit. For $\delta=15$~keV the DAMA region
includes the best fit point of the CoGeNT modulation, hence in this case the
two modulation signals are in excellent agreement. However, the
corresponding region is excluded by essentially all limits shown in the
plot, including the one from CoGeNT (unmod). If the DM mass splitting is
increased to 25~keV, the CoGeNT modulation best fit point moves to smaller
DM masses, becoming consistent with CDMS, SIMPLE, and CoGeNT (unmod).
However, in this case the agreement between DAMA and CoGeNT (mod) is lost
and the XENON100 limit becomes severe, since the heavy target is favored for
inelastic scattering. Hence, a situation similar to inelastic scattering for
$f_n = f_p$ as shown in fig.~\ref{fig:cogent-iSI} and discussed in
sec.~\ref{sec:inel} is obtained.


\section{Spin-dependent interactions}
\label{sec:SD}

Let us briefly mention also the situation in the case of spin-dependent
(SD) interactions. Here the relative importance of the various experiments
is largely changed compared to the SI case, since the $A^2$ factor (or
$A^2_{\rm eff}$ in the case of general isospin dependence) is replaced by
the corresponding nuclear structure function taking into account the
spin-structure of the nuclei, see eq.~\eqref{eq:csSD}. Depending on whether
the spin of a nucleus is carried mainly by neutrons or protons there is a
strong dependence on the isospin coupling, parametrized by $\tan\theta =
a_n/a_p$ in eq.~\eqref{eq:thetaSD}.

\begin{figure}
  \includegraphics[height=0.3\textwidth]{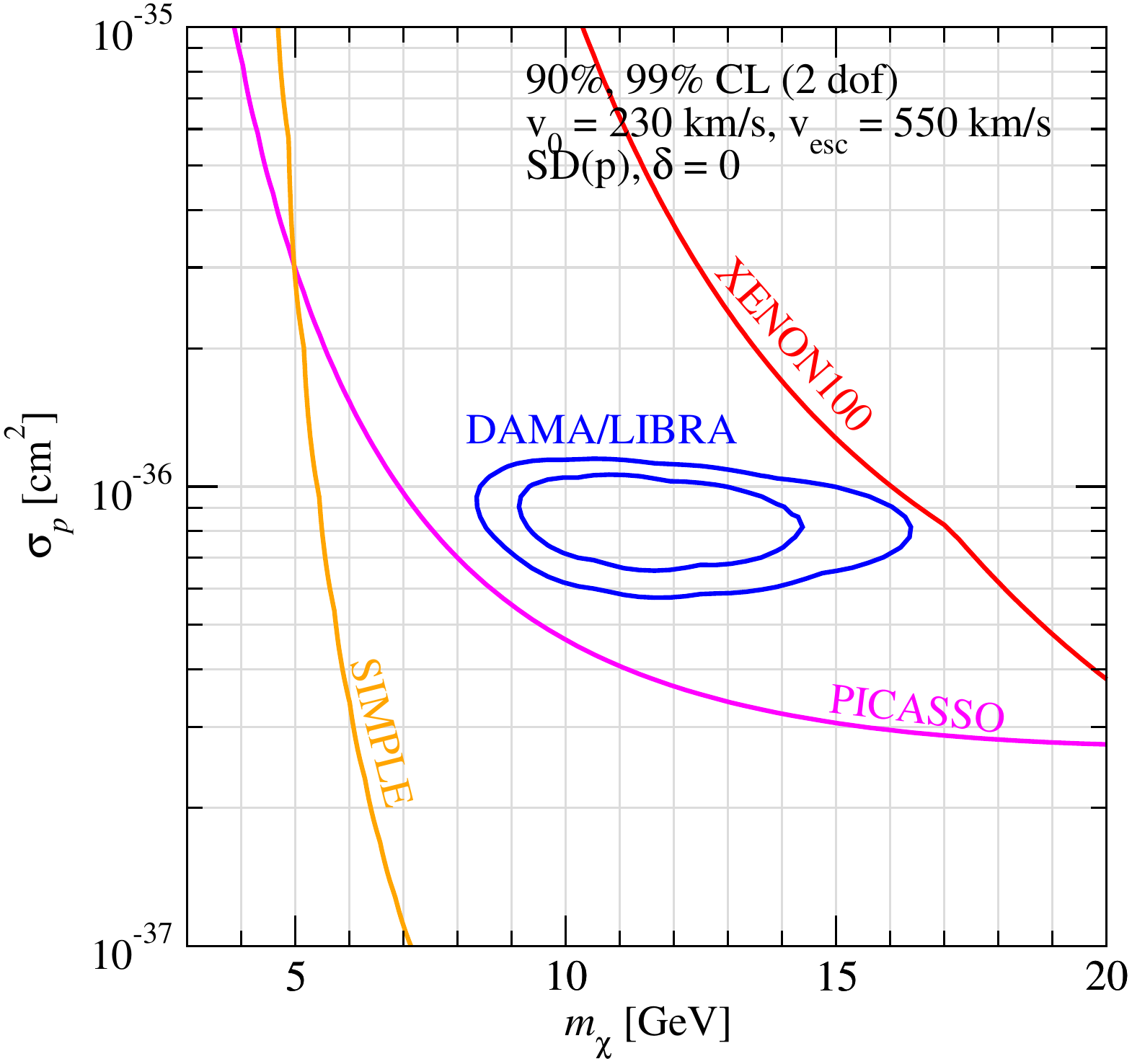} 
  \includegraphics[height=0.3\textwidth]{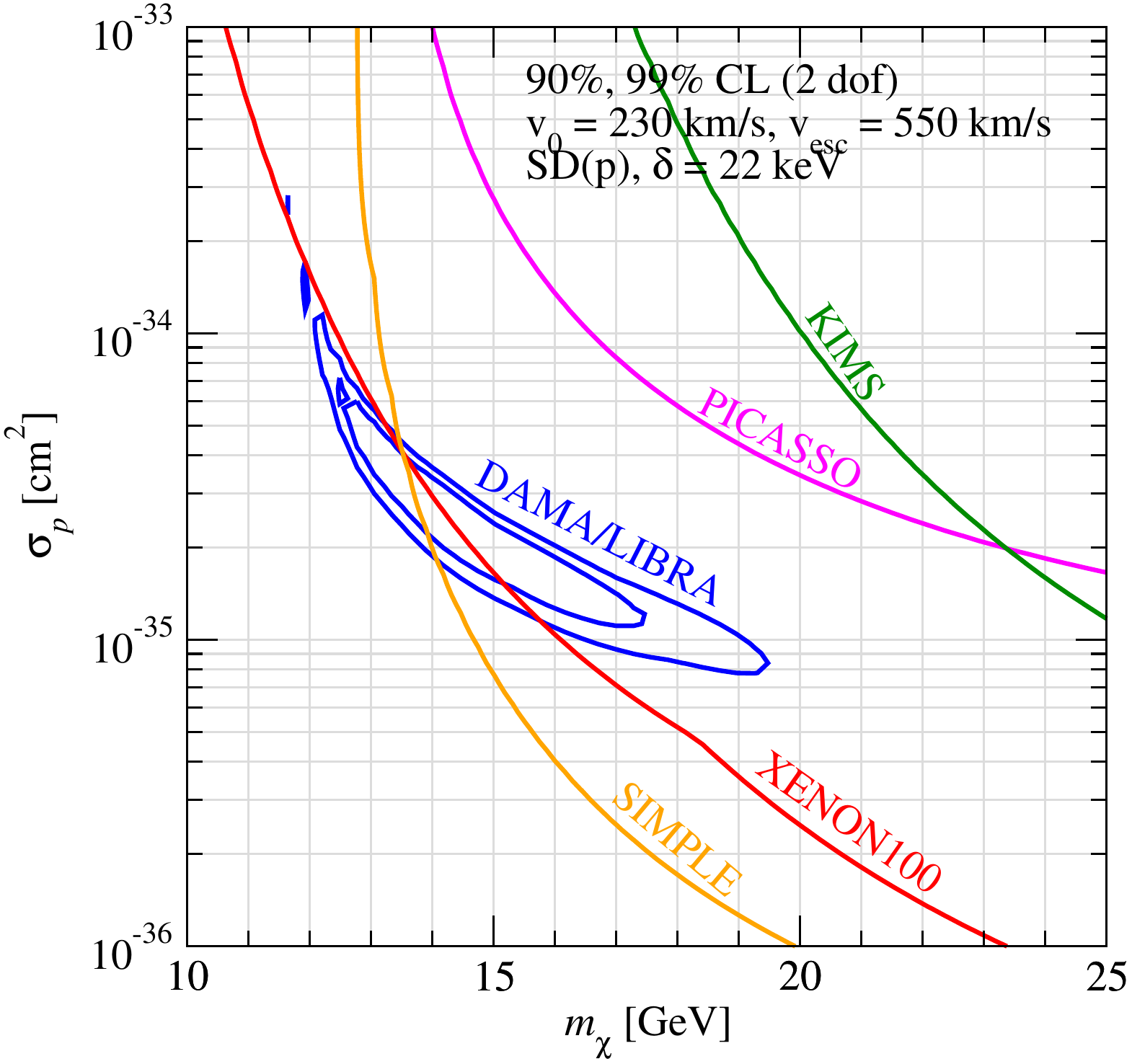} 
  \includegraphics[height=0.3\textwidth]{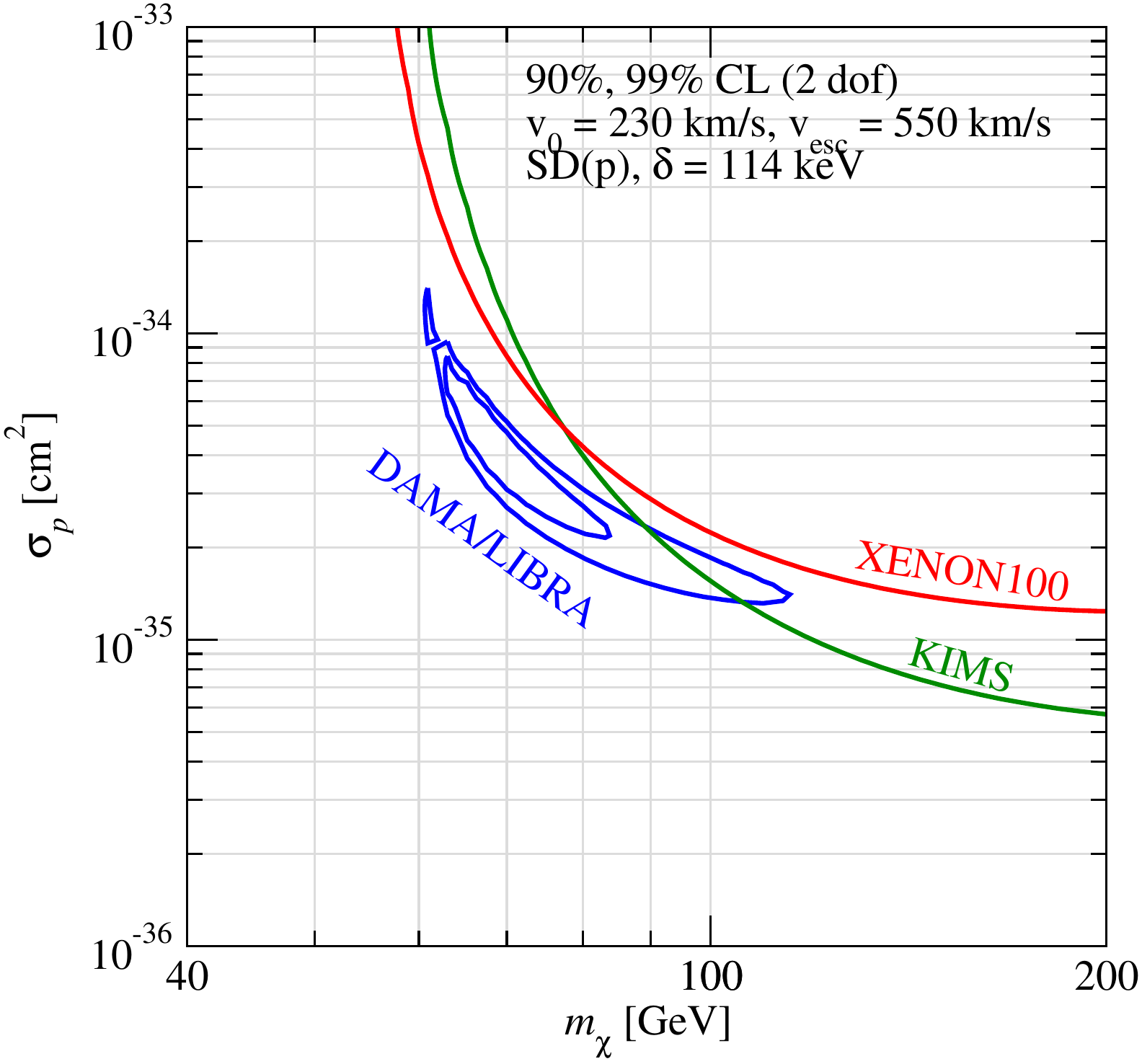}
  \mycaption{Allowed regions for DAMA/LIBRA~\cite{Bernabei:2008yi} at 90\%,
  99\%~CL and limits from XENON100~\cite{Aprile:2011hi},
  SIMPLE~\cite{Felizardo:2011uw}, PICASSO~\cite{Archambault:2009sm}, KIMS~\cite{Lee:2007qn}
  at 90\%~CL in the DM
  mass--cross section plane for spin-dependent scattering off protons
  ($\tan\theta = 0$ in eq.~\ref{eq:thetaSD}). The left panel corresponds to
  elastic scattering, whereas the middle and the right panel correspond to
  inelastic scattering with DM mass splittings $\delta = 22$~keV and $\delta
  = 114$~keV, respectively.}
  \label{fig:SD}
\end{figure}  

The case of coupling only to the spin of the proton ($\theta=0$) is
especially interesting for DAMA, since large part of the spin of Na and I
are carried by protons. In this case, however, there are important bounds from experiments that use
fluorine as a target: from COUPP~\cite{Behnke:2008zza},
PICASSO~\cite{Archambault:2009sm}, and most recently from
SIMPLE~\cite{Felizardo:2011uw}. As shown in the left panel of
fig.~\ref{fig:SD} the fluorine bounds exclude the DAMA region. One possibility
to make the DAMA explanation consistent with the bounds from the other experiments is to assume that the scattering is both spin dependent and inelastic, since then the rate for scattering on the light
fluorine target is reduced. This possibility was discussed in terms of effective DM tensor interactions in Ref.~\cite{Kopp:2009qt} and 
in Ref.~\cite{Chang:2010en} by assuming that DM carries a large  coupling to magnetic nuclear moments.
As shown in the middle and right panels of fig.~\ref{fig:SD}
for DM mass splittings of $\delta =22$~keV and $\delta = 114$~keV
unconstrained allowed regions for DAMA appear.\footnote{Quantitative
differences compared to the results of \cite{Kopp:2009qt} appear due to
slightly different values adopted for the halo parameters $v_0$ and $v_{\rm
esc}$.} These two solutions correspond to scattering off Na and I in DAMA,
respectively. For large mass splittings the bound from the KIMS experiment
(using a CsI target) become more relevant, but are still not able to exclude
the DAMA region. Other constraints such as from Ge experiments or from the
CRESST tungsten data disappear, since for those nuclei protons do not
contribute to the spin. This, however, at the same time makes an explanation of CoGeNT data
in this framework impossible. Constraints from neutrinos from DM
annihilations in the sun have been considered for these DAMA solutions in
\cite{Shu:2010ta}.

\begin{figure}
  \includegraphics[width=0.5\textwidth]{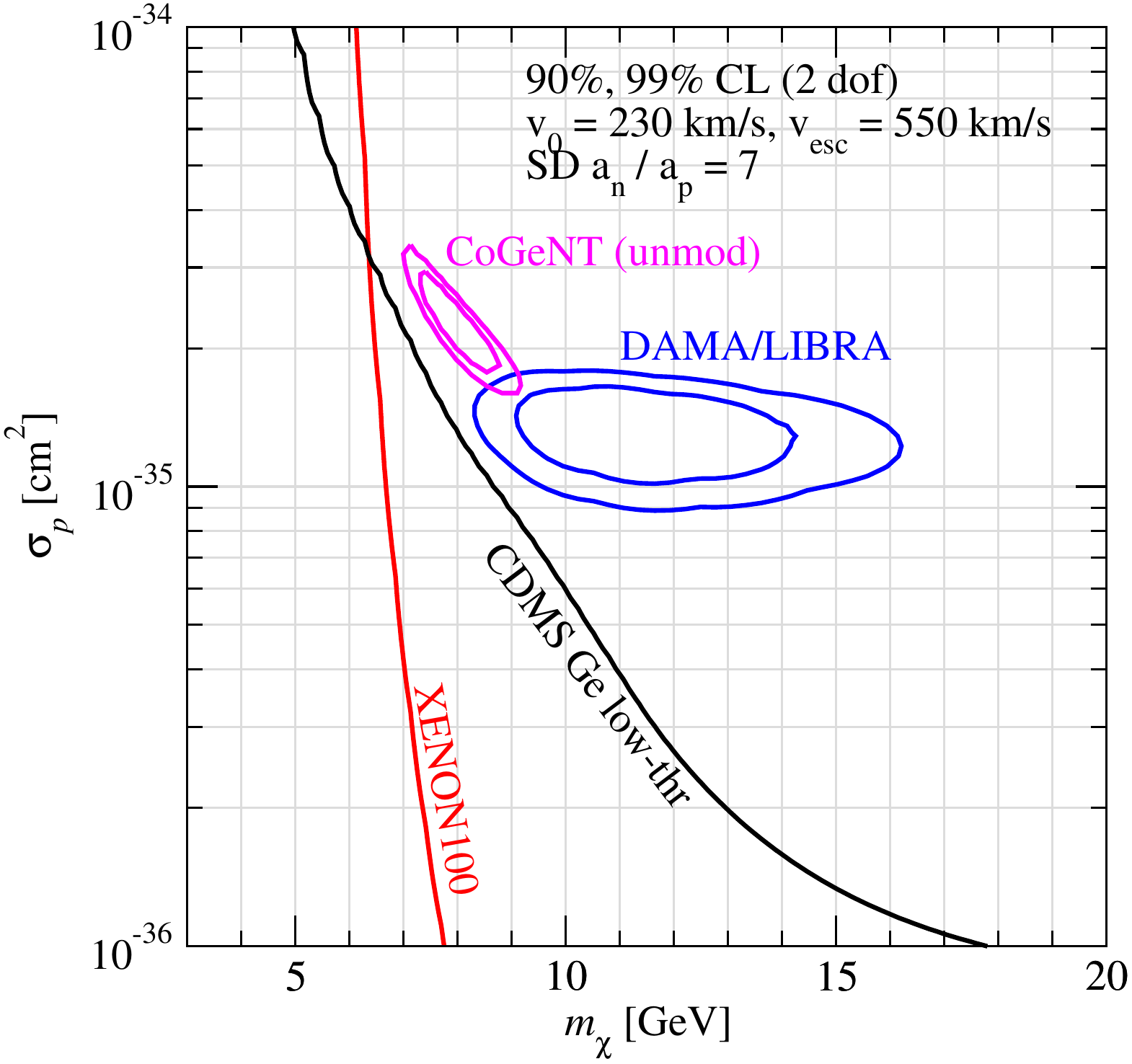}
  \mycaption{Allowed regions (90\% and 99\%~CL) and limits (90\%~CL) in the
  DM mass--cross section plane for elastic spin-dependent scattering with 
  dominant coupling to neutrons ($a_n/a_p \approx 7$). We
  use data from CoGeNT~\cite{Aalseth:2011wp} (unmodulated),
  DAMA/LIBRA~\cite{Bernabei:2008yi} modulation, CDMS low-threshold Ge
  \cite{Ahmed:2010wy}, and XENON100~\cite{Aprile:2011hi}.} 
  \label{fig:SD-neutron}
\end{figure}

For the germanium target in CoGeNT the spin is dominated by neutrons. However,
in this case the constraints from DM scattering on xenon are particularly severe. By scanning
the parameter $\theta$ we have found best compatibility of CoGeNT and DAMA
for $\theta = 0.445 \pi$, where $a_n/a_p \approx 7$, i.e., large coupling to
neutrons. As visible in fig.~\ref{fig:SD-neutron}, in this case the 99\%~CL 
regions of DAMA and CoGeNT (unmod)
overlap. The corresponding region is excluded by CDMS low-threshold
Ge data (which obviously follows the same pattern as CoGeNT) but also by the
XENON100, due to the large contribution to the spin of the $^{129}$Xe and
$^{131}$Xe isotopes from the neutron. 

\section{Light mediators}
\label{sec:mediators}

Up until now we have always assumed a contact interaction between DM and quarks, resulting from integrating out heavy mediator particle(s). ``Heavy'' means here masses large compared to the transferred momentum, i.e., $m_\phi^2 \gg \bs{q}^2 = 2m_A E_d \sim (100\,\rm MeV)^2$. For masses smaller than this a nontrivial $\bs{q}^2$ dependence will appear in the differential cross section. As a proxy we use the $\bs{q}^2$ dependence that results from an exchange of light scalars, see eq.~\eqref{eq:cs2}. In particular, for a "massless" mediator with $m_\phi^2 \ll \bs{q}^2$ an additional factor $1/E_d^2$ will appear, leading to an enhancement of the rate at low recoil energies. In this section we investigate whether light mediators could improve some of the fits discussed above due to the additional $1/\bs{q}^2$ enhancement at low recoil energies. This could also have an additional benefit. Many of the considered models, such as inelastic scattering or neutron/proton cancellations, lead to a suppressed rate on the nucleus level, which has to be compensated by large DM--nucleon cross sections. One possible way to achieve such large cross sections is to consider light mediator particles. 

\begin{figure}
  \includegraphics[width=0.5\textwidth]{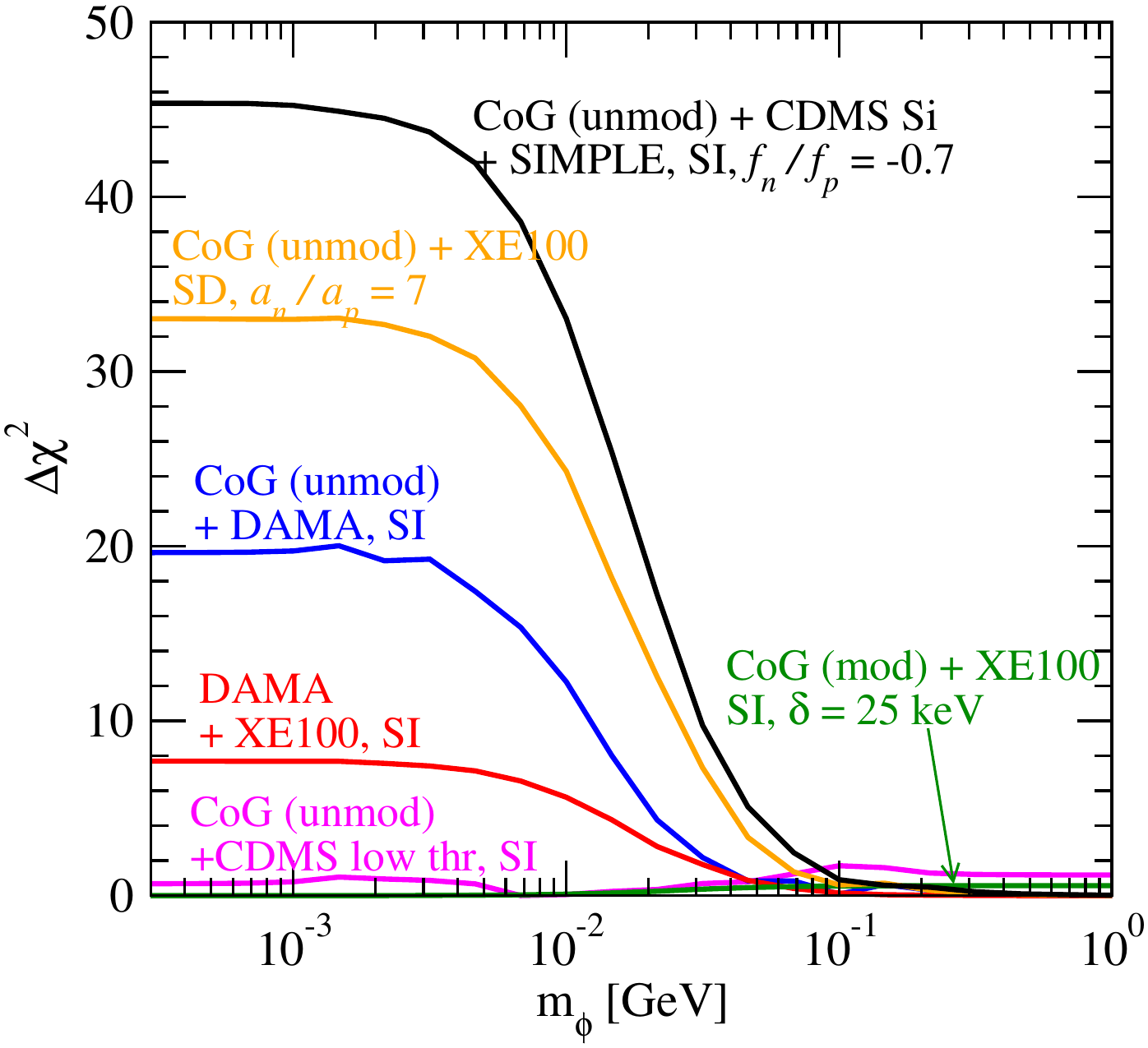}
  \mycaption{$\Delta\chi^2$ as a function of the mediator mass for various combinations of conflicting data sets within various DM interaction models. } 
  \label{fig:mediator}
\end{figure}  

In fig.~\ref{fig:mediator} we show the effect of decreasing the mediator mass for various combinations of conflicting data sets of the previous sections. For each combination we show the 
$\Delta\chi^2$ with respect to the $\chi^2$ minimum. Since we combine conflicting data sets (with the aim of improving the fit by reducing the mediator mass) the absolute value of the $\chi^2$ is not of interest to us here. As is clear from the figure the fit does not improve for light mediators---on the contrary, in many cases the fit gets considerably worse. 

This is the case, for example, for the CoGeNT unmodulated excess versus DAMA (blue curve) and for DAMA excess versus XENON100 (red curve) assuming spin-independent DM scattering (cf. also fig.~\ref{fig:regions-all-eSI}). Similarly, the conflict between the unmodulated CoGeNT rate versus the CDMS Si and SIMPLE constraints in the case of spin-independent scattering with $f_n/f_p = -0.7$ found in fig.~\ref{fig:isospin} (black curve) becomes significantly worse for $m_\phi \lesssim 100$~MeV. The same conclusion applies for the conflict between the CoGeNT rate versus the XENON100 bound in the case of spin-dependent interactions mainly on neutrons found in fig.~\ref{fig:SD-neutron}  (orange curve). In all of these cases actually a lower bound on the mediator mass can be obtained, if the fit for large $m_\phi$ is accepted despite the tension between the data sets. 
The light mediator also does not remove the conflict between CoGeNT unmodulated spectrum versus CDMS low-threshold Ge data for SI scattering (cf. fig.~\ref{fig:regions-all-eSI}), though in this case the $\chi^2$ remain basically flat compared to contact interaction. The same conclusion applies to the tension between CoGeNT modulation data and XENON100 bounds on inelastic scattering (cf.~fig.~\ref{fig:cogent-iSI}), where again light mediators provide a fit of similar quality as the contact interaction. 

These results seem to be in disagreement with the ones of \cite{Foot:2011pi}, where DM interacts via a massless mediator (in that case the photon) with the nuclei, and a consistent fit for CoGeNT, DAMA, and XENON100 is obtained. However, no direct comparison with our results is possible, since in Ref.~\cite{Foot:2011pi} a multi-component DM model is considered with very peculiar halo properties due to strong self-interactions of the DM.

\section{Conclusions}
\label{sec:conclusions}

In conclusions, we have investigated the recent indication of annual modulation in the CoGeNT data, adopting the working hypothesis that this is a signal of dark matter scattering on nuclei. Assuming elastic spin-independent scattering, we find that there is tension between the modulation signal and the time integrated event excess in CoGeNT. For DM parameter values consistent with the unmodulated rate, the modulation amplitude in the energy range of 0.9--3~keV (where the significance of the modulation signal is highest) is too small to explain the data. Furthermore, we do not find 
any region of parameters where both CoGeNT and DAMA can be explained without being in conflict with bounds from XENON100 and CDMS. Motivated by these problems we considered a variety of more exotic DM interaction hypotheses.
We tested inelastic spin-independent DM scattering, as well as elastic and inelastic spin-dependent DM scattering, allowing for isospin breaking interactions and light mediators. In all of the cases we find that it is not possible to reconcile both CoGeNT and DAMA at the same time, while also obeying existing bounds from other direct detection experiments. However, part of the data can be made consistent in some cases. For example, in a small parameter region the modulation signal in CoGeNT 
becomes consistent with the unmodulated rate (but cannot explain it, nor the DAMA signal), as well as with other bounds at 90\%~CL for inelastic spin-independent scattering with $\delta \sim 25$~keV and isospin breaking tuned to avoid xenon constraints \cite{Frandsen:2011ts} (cf.\ fig.~\ref{fig:isospin-inel}). The DAMA signal (but not CoGeNT) can be due to DM if the scattering is both spin dependent and inelastic \cite{Kopp:2009qt} (cf.\ fig.~\ref{fig:SD}).

As a general comment we mention that both the case of inelastic scattering as well as the case of cancellations between scatterings on
neutrons and protons  lead to a suppression of the event rate, which
requires very large cross sections to account for possible signals in
DAMA/CoGeNT, typically larger than $10^{-39}\,{\rm cm}^2$. It remains to be
shown that such large cross sections can be consistent with constraints from
neutrinos~\cite{Shu:2010ta,Chen:2011vd} and colliders~\cite{Goodman:2010ku,Goodman:2010yf,Fortin:2011hv,Fox:2011fx,Bai:2010hh,Mambrini:2011pw}. 

Finally let us comment on preliminary results from the CRESST-II experiment, showing an unexplained event excess in their O-band data. They reported 32 observed events with an expected background of $8.7\pm 1.4$ \cite{seidel_iDM10}. A crude estimate indicates that such events could be consistent with elastic spin-independent 
scattering with similar parameter values as CoGeNT and DAMA \cite{Schwetz:2010gv}. For example, for $m_\chi = 10$~GeV and $\sigma_p = 10^{-40}\, \rm cm^2$ the predicted number of oxygen scatters for 400~kg~day exposure with 100\% efficiency is about 16 events, taking into account the reported individual thresholds of the 9 detectors. However, the indicated parameter region 
 is again  excluded by XENON100 and CDMS bounds \cite{Schwetz:2010gv}. For the model considered in fig.~\ref{fig:isospin-inel} we expect about 50 events for $m_\chi = 12$~GeV, $\sigma_p = 5\times 10^{-37}\, \rm cm^2$, and $\delta = 15$~keV, whereas for
$\delta = 25$~keV no scattering is possible on oxygen with CRESST-II thresholds due to kinematics. Similarly, also the spin-dependent inelastic scattering able to explain DAMA cannot accommodate the potential signal in CRESST, since $^{16}$O has no spin, and the inelasticity disfavors the scattering kinematically.

\acknowledgments
We thank Juan Collar for providing us with the CoGeNT raw data, which we used for cross checking the data used in this analysis. We would like to thank Felix Kahlhoefer, Joachim Kopp, and Gilles Gerbier for useful discussions. 
The work of T.S.\ was partly supported by the Transregio Sonderforschungsbereich TR27
``Neutrinos and Beyond'' der Deutschen Forschungsgemeinschaft.

\bibliographystyle{my-h-physrev.bst}
\bibliography{./cog}

\end{document}